\documentclass{aa}
\usepackage{graphicx}
\usepackage{txfonts}
\usepackage{booktabs}
\usepackage{natbib}
\usepackage[colorlinks=true, urlcolor=blue, pdfencoding=auto, linkbordercolor={1 0 0}]{hyperref}

\usepackage{color} 

\hypersetup{
    colorlinks=true, 
    linkcolor=blue,  
    citecolor=blue,  
    filecolor=magenta, 
    urlcolor=blue,    
}

\begin{document}

   \title{The invariance of group occupation across the cosmic web}

   \author{
   Vicente Izzo Dominguez\inst{1} \and 
   Facundo Rodriguez\inst{2, 3} \and
   Antonio D. Montero-Dorta\inst{1} }

   \institute{
   Departamento de F\'isica, Universidad T\'ecnica Federico Santa Mar\'ia, Avenida Vicu\~na Mackenna 3939, San Joaqu\'in, Santiago, Chile.
   \and
   CONICET. Instituto de Astronomía Teórica y Experimental (IATE). Laprida 854, Córdoba X5000BGR, Argentina.
   \and
   Universidad Nacional de Córdoba (UNC). Observatorio Astronómico de Córdoba (OAC). Laprida 854, Córdoba X5000BGR, Argentina.
   }

   \date{Received -; accepted -}

 
    \abstract
    {Recent efforts to identify secondary variations in the Halo Occupation Distribution (HOD) have primarily focused on simulations examining the role of large-scale cosmic environments, such as superclusters, filaments, and underdense regions or voids. If present, these variations could yield valuable insights into galaxy formation mechanisms, halo assembly processes, and the influence of external factors on the evolution of cosmic structure.}
    {We aim to explore whether the secondary trends in the HOD driven by the large-scale structure of the Universe are present observationally. In particular, we examine whether the HOD depends on the distance to key features of the cosmic web by explicitly quantifying these spatial relationships. We further analyze whether HODs vary across different cosmic environments, as defined by critical point classifications, and assess the influence of intrinsic galaxy properties, such as central galaxy color.}
    {We create volume-limited galaxy samples from SDSS-DR18 and use a group catalog to determine halo masses, as well as identify central and satellite galaxy membership. Additionally, we employ a DisPerSE catalog to locate critical points such as maxima, minima, and filaments in the cosmic web. We assess how the halo occupation distribution (HOD) varies based on proximity to these features and analyze these variations across five distinct cosmic environments. Furthermore, we investigate trends related to the color of central galaxies and test the reliability of our results by using alternative DisPerSE catalogs generated with different smoothing scales and persistence thresholds.}
    {Our analysis confirms that the large-scale cosmic environment only weakly influences the HOD. However, second-order environmental dependencies might be revealed through a multivariate approach combining both local and large-scale environment metrics with intrinsic galaxy properties. Future investigations employing next-generation surveys with improved statistical power, coupled with sophisticated modeling techniques, may provide the necessary precision to detect and characterize these subtle environmental correlations.}
    {}


   \maketitle
%
The formation and evolution of galaxies within the cosmic web represents one of the most complex challenges in modern astrophysics. According to the standard paradigm of hierarchical structure formation, galaxies form and evolve within the potential wells of dark matter halos \citep{WhiteRees1978}. While the growth of these halos is governed primarily by gravitational processes such as accretion and mergers \citep{Springel2005, Wechsler_2018}, the resulting galaxy populations are shaped by a rich interplay of baryonic physics including gas cooling, star formation, and various feedback mechanisms \citep{Baugh_2006, Somerville_2015, Naab_2017}. The different timescales involved in halo mergers compared to galaxy mergers further complicate this picture, as galaxies may retain their identity long after their host halos have merged, making galaxy evolution sensitive to both the local baryonic processes and the larger-scale dynamical history of the halo environment \citep{Lacey_1994, Hopkins2008}.

The Halo Occupation Distribution (HOD) framework has proven to be one of the most successful approaches for quantifying the relationship between galaxies and their host halos. By characterizing the statistical distribution of galaxies within dark matter halos as a function of halo mass \citep{Berlind_2002, COORAY_2002}, the HOD provides a powerful tool for connecting theoretical predictions with observational data. This methodology has found numerous applications across astrophysical research. It has been used extensively to interpret galaxy clustering measurements, providing insights into how different galaxy populations trace the underlying matter distribution \citep{Zheng_2005, Zehavi2011}. The HOD framework has enabled detailed studies of galaxy demographics, revealing how various galaxy types occupy halos of different masses across redshift  (e.g., \citealt{Tinker_2008, Contreras2017, Contreras2023}). Furthermore, it has become an essential component in the construction of realistic mock galaxy catalogs needed for survey planning and analysis \citep{Behroozi2019}. Perhaps most significantly, the HOD has emerged as a valuable tool for cosmological studies, offering constraints on fundamental parameters through its connection to galaxy bias and large-scale structure \citep{McCracken2015, Avila2020}. The widespread adoption of the HOD approach stems from its unique combination of physical insight, flexibility, and computational efficiency \citep{Kravtsov_2004, vandenBosch2013}.

Recent advances in our understanding of galaxy formation have revealed that the HOD depends not only on halo mass but also on the larger-scale environment in which halos reside \citep{Zentner2014, Salcedo2020}. Significant variations appear in regions of extreme density contrast. In the underdense environments of cosmic voids, halos consistently show reduced occupation numbers compared to field halos of similar mass, along with characteristically later formation times \citep{Tinker_2006, Tinker_2009, Alfaro2020, Alfaro_2022}. Conversely, in the overdense environments or future virialized structures (FVSs), halos exhibit enhanced occupation numbers and accelerated evolutionary timelines \citep{Aragon-Calvo2010, Cautun2014, Alfaro2021, Alfaro_2022}. These environmental dependencies reflect fundamental differences in halo assembly histories and the resulting galaxy formation efficiency \citep{Ricciardelli2014, Pollina2017, Alfaro_2020}. The environment and the geometry of the tidal field around halos have also been invoked to explain secondary halo bias effects such as halo assembly bias\footnote{This effect, which has only been measured in numerical simulations, refers to the dependence of halo clustering or bias on halo  accretion history and related quantities, at fixed halo mass.} (e.g.,  \citealt{ShethTormen2004,Gao2005,Dalal2008, Borzyszkowski2017, Musso2018, Paranjape2018, SatoPolito2019, Ramakrishnan2019, Contreras2019, Ramakrishnan2020, Tucci2021, Contreras2021_cosmo, Balaguera2024, MonteroDorta2025}), which might also impact galaxy clustering and the HODs (e.g., \citealt{Zehavi2018, Artale2018, MonteroDorta2020, MonteroDorta2021, Wang2022, montero_facu_2024, Alam2024,Kim2025}).

The cosmic web --  the intricate network of voids, filaments, walls, and nodes that constitutes the large-scale structure of the Universe -- provides the essential environmental context for understanding these variations in galaxy occupation \citep{Bond1996, Libeskind2017}. Voids, which dominate the volume of the Universe, represent regions of exceptionally low density where galaxy formation is strongly suppressed due to both the lack of material and the expansion-dominated dynamics of these regions \citep{Ceccarelli2013, Hamaus2014}. Filaments form the connecting bridges between these voids, channelling matter flows toward the denser nodes and influencing the properties of galaxies along these pathways \citep{Tempel2014, Malavasi2017}. The nodes, where multiple filaments intersect, correspond to the densest regions of the cosmic web and typically host massive galaxy clusters \citep{Aragon-Calvo2010, Hahn_2007}. This environmental diversity across the cosmic web drives the observed heterogeneity in galaxy populations properties and their occupation statistics \citep{Dekel_2009, Kraljic2018, rodriguez2025centralgalaxyalignmentsdependence, alfaro2025evolutionhalooccupationdistribution}, making the environmental context an essential consideration for any complete understanding of galaxy formation.

Using Illustris-TNG hydrodynamical simulations (\citealt{Pillepich2018b,Pillepich2018,Springel2018}), \cite{Perez2024} recently studied how the cosmic web affects the HOD. This research focuses on analysing how HODs change in halos found in filaments and nodes. The findings reveal a strong correlation between the HOD of halos in filaments and the overall galaxy sample, suggesting that halo mass is the most significant factor affecting galaxy distribution. In contrast, nodes exhibit an excess of faint galaxies in lower-mass halos and a bimodal color distribution among low-mass galaxies, indicating signs of early reddening.

In this work, we investigate potential secondary dependencies of the HOD on the large-scale cosmic environment, using spectroscopic data from the Sloan Digital Sky Survey (SDSS; \citealt{York_2000}). Specifically, we aim to determine whether the number of galaxies within halos varies systematically with proximity to prominent features of the cosmic web and across distinct environmental classifications. To this end, we combine group catalog data obtained through a halo-based group identification algorithm \citep{Rodriguez2020} with a cosmic web classification based on the DisPerSE algorithm \citep{Malavasi2020}. This analysis is particularly relevant as a means of constraining the halo-galaxy connection from observations.

The remainder of this paper is structured to systematically explore the relationship between galaxy occupation statistics and cosmic environment. In Sect.~\ref{sec:data}, we present the observational datasets forming the basis of our analysis, with particular focus on the cosmic web critical point sample and its distinctive features. Section~\ref{sec:environment} details our methodology for cosmic environment classification, explaining the quantitative criteria distinguishing voids, filaments, and nodes within the large-scale structure. The core of our analysis appears in Sect.~\ref{sec:results}, where we present a comprehensive examination of the environmental dependence of halo occupation statistics, including the systematic variations across different cosmic web environments and their statistical significance. Finally, Sect.~\ref{sec:conclusions} synthesizes our principal findings, discussing their implications for galaxy formation theory and presenting promising directions for future research in this field.

\section{Data}
\label{sec:data}
In this section, we provide a detailed description of the three catalogs utilized in this study: the galaxy catalog, the galaxy groups catalog, and the DisPerSE critical points catalog.

\subsection{SDSS Galaxy Sample}

The Sloan Digital Sky Survey (SDSS) is one of the most comprehensive astronomical surveys ever conducted, providing a detailed map of the sky with a focus on understanding the large-scale structure of the universe \citep{York_2000}. Since its inception, SDSS has delivered high-quality photometric and spectroscopic data for millions of celestial objects, enabling a wide range of astrophysical studies. Observations are carried out using a 2.5-meter telescope at Apache Point Observatory in New Mexico, USA. The telescope is equipped with a wide-field imaging camera and a spectrograph that covers five optical bandpasses (u, g, r, i, z). The spectrograph is capable of obtaining up to 1000 spectra simultaneously over a wavelength range of 3800–9200 $\AA$ with a resolution of $\lambda/\Delta\lambda \sim 2000$ \citep{maraston2009}.

The 18th data release (DR18) of the Sloan Digital Sky Survey (SDSS) marks the first release for SDSS-V, the fifth generation of the survey. SDSS-V comprises three primary scientific programs: the Milky Way Mapper (MWM), the Black Hole Mapper (BHM), and the Local Volume Mapper (LVM) \citep{almeida2023}. DR18 includes extensive targeting information for the multi-object spectroscopy programs (MWM and BHM), providing input catalogs and selection functions for a wide array of scientific objectives. It also incorporates all previous SDSS data releases, covering more than 14,000 deg$^2$ of the sky and offering spectroscopic data for millions of galaxies, stars, and quasars. Furthermore, DR18 adds approximately 25,000 new SDSS spectra, along with supplemental information for X-ray sources identified by eROSITA in its eFEDS field.

For our analysis, we used the SDSS-DR18 Main Galaxy Sample (MGS), which provides a robust dataset for studying the large-scale distribution of galaxies. We focused on galaxies within the Legacy footprint area, applying magnitude and redshift constraints to construct a volume-limited sample suitable for tracing cosmic structures. Specifically, we selected galaxies with apparent r-band magnitudes brighter than 17.77 and redshifts below 0.3, consistent with previous studies using SDSS data \citep{Strauss_2002}. To ensure uniform completeness across the sample, we imposed a redshift limit of $z_{\mathrm{lim}} = 0.1$ and a corresponding absolute magnitude threshold of $M_{\mathrm{r}} = -19.77$. This selection yields a well-defined and nearly uniform sample across the surveyed area, ideal for identifying filaments, nodes, and voids.

\subsection{Galaxy Group Catalog}
\label{subsec:groups}

Galaxy groups are observational tracers of dark matter halos and are fundamental for exploring the galaxy–halo connection. In this work, we use a group catalog from SDSS DR18, constructed using the method developed by \citet{Rodriguez2020}. This approach combines a Friends-of-Friends (FoF) algorithm \citep{Huchra1982,Merchan_2005} with a halo-based refinement procedure \citep{Yang_2007}, enabling the identification of gravitationally bound systems and the assignment of physical halo properties.

The FoF algorithm initially links galaxies into candidate groups based on their projected separation and line-of-sight velocity differences. These preliminary associations are then refined iteratively using an abundance matching technique \citep{Vale2004,Kravtsov_2004,Behroozi_2010}, which estimates halo masses by assuming a monotonic relationship between total group luminosity and halo mass. This iterative process continues until convergence is reached in both group membership and halo mass estimates.

The resulting catalog spans a broad range of systems—from small groups to rich clusters—and provides key properties such as galaxy membership, halo mass ($M_{\text{group}}$), and spatial coordinates. Halo mass estimates from this catalog are in good agreement with independent measurements from weak gravitational lensing \citep{gonzalez2021}, and the group properties have been shown to reproduce the observed central–satellite distributions seen in simulations \citep{Alfaro_2022, 10.1093/mnras/stad623}.

In this framework, the brightest galaxy in each group is classified as the central galaxy, while the remaining members are considered satellites. For this study, we restrict our analysis to groups with at least one spectroscopic member to ensure reliable estimates of group properties and robust measurements of the HOD.

\subsection{DisPerSE Critical Points Sample}
\label{subsec:disperse}

The cosmic web filaments and critical points in this study were identified using DisPerSE \citep{sousbieetal2011, sousbie2011}, which applies discrete Morse theory to 3D galaxy distributions. The algorithm first reconstructs the density field using the Delaunay Tessellation Field Estimator (DTFE) method \citep{schaap2000continuousfieldsdiscretesamples, 2009LNP...665..291V}, creating a tetrahedral mesh from galaxy positions and computing densities through local averaging. We considered three smoothing levels: unsmoothed (SD0), single (SD1), and double smoothing (SD2).

Critical points (maxima, minima, and saddles) are located where the density gradient vanishes, with filaments traced as field lines connecting maxima to saddles. Persistence filtering at $3\sigma$ and $5\sigma$ thresholds \citep{sousbie2011, 2017MNRAS.470.1274M} removes noise-induced features, where $5\sigma$ retains only the most significant structures. 

The identification of cosmic web structures in this study is based on the filament catalogs generated with the DisPerSE algorithm and made publicly available by \citet{Malavasi2020}. In their work, DisPerSE was applied to the Main Galaxy Sample of SDSS DR12, producing multiple catalogs of critical points and filaments by varying both the smoothing scale of the galaxy density field and the persistence threshold that controls the significance of topological features. These catalogs provide a detailed and flexible mapping of the cosmic web, including the positions of nodes, saddles, and filamentary structures. The datasets used in this analysis were obtained from an online repository\footnote{\url{https://l3s.osups.universite-paris-saclay.fr/cosfil.html}} and serve as the basis for our characterization of the large-scale environment. The robustness of the DisPerSE method and the systematic exploration of its parameters make it particularly well-suited for studies of the cosmic web.

\begin{figure}
    \centering
    \includegraphics[width=\linewidth]{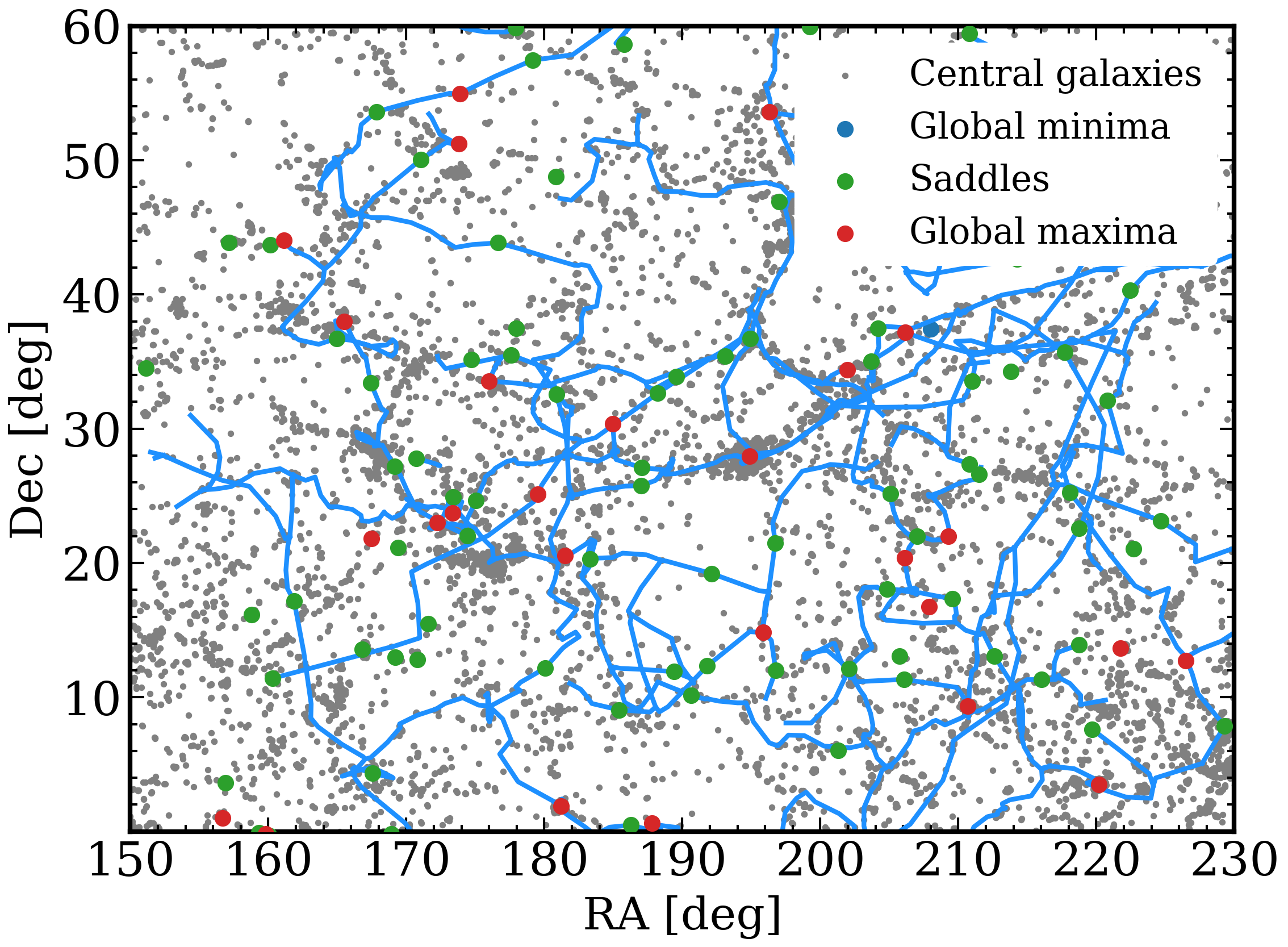}
    \caption{Galaxy and critical point distribution in a $\pm 75$ Mpc redshift slice centered on the Coma cluster. Grey points are central galaxies from the SDSS MGS sample. Large circles are DisPerSE critical points color-coded by type (nodes, voids, saddles). A one-smoothing of the density field was applied before structure identification.}
    \label{fig:cp match}
\end{figure}

To associate galaxy groups with these cosmic web structures, we followed the matching procedure outlined in the same work. This involves converting galaxy positions from celestial coordinates to comoving Cartesian space, ensuring compatibility with the coordinate system of the filament catalogs and enabling a consistent spatial comparison between galaxy groups and cosmic web features. The spatial distribution of critical points (nodes, voids, and saddles) is shown in Fig.~\ref{fig:cp match}, in a redshift slice centered on the Coma cluster.

\section{Cosmic Environment Classification}
\label{sec:environment}

\begin{figure*}[t]
    \centering
    \includegraphics[width=1\linewidth]{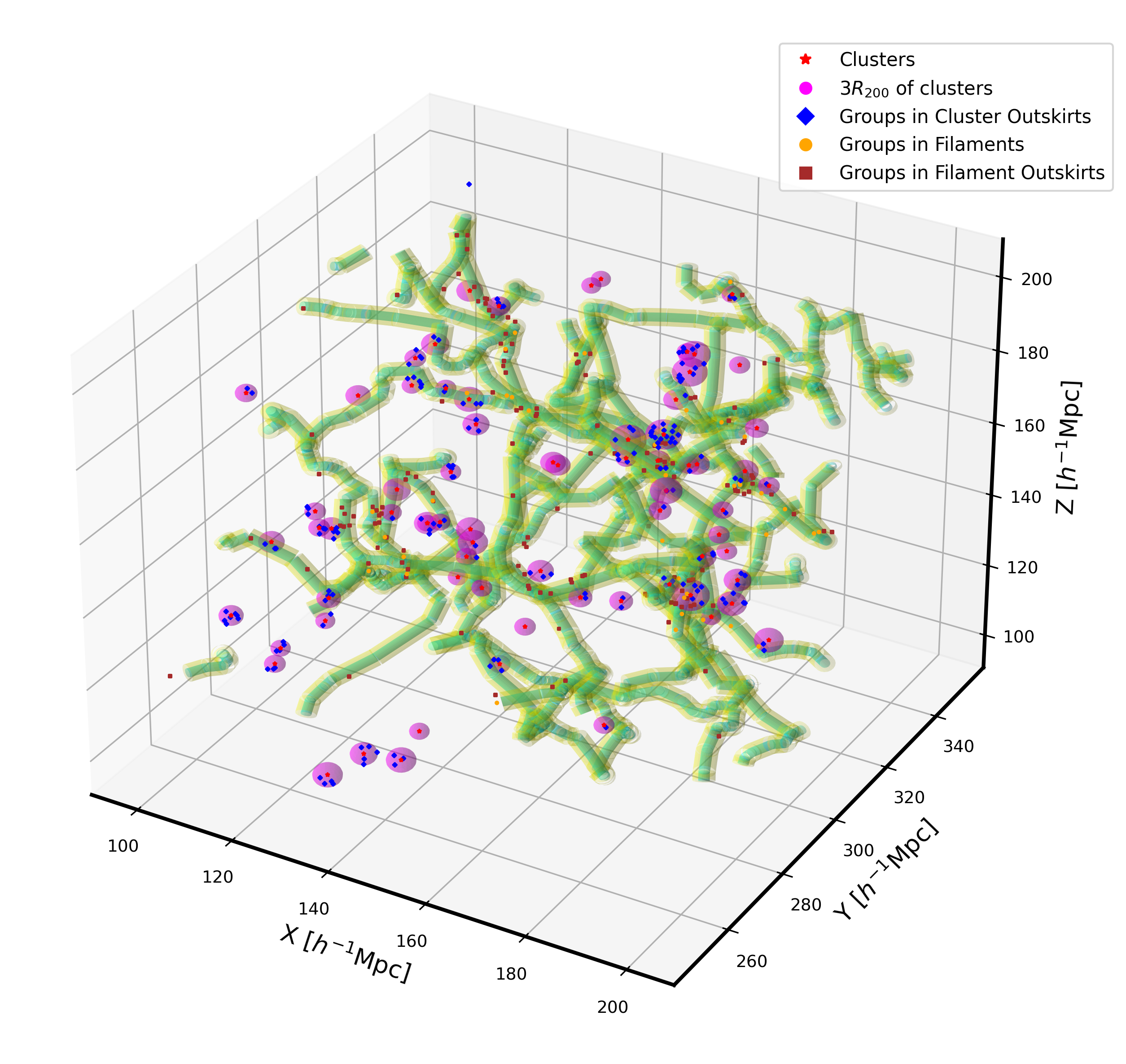}
    \caption{
Three-dimensional visualization of the five cosmic environments used in our analysis: galaxy clusters (red), cluster outskirts (blue), filaments (yellow), filament outskirts (brown), and other environments (not directly associated with these structures). Filamentary structures were extracted using the DisPerSE algorithm. Groups are shown as transparent magenta spheres, with radii equal to \(3 \times R_{200}\) of each halo. Galaxies classified as Filaments lie within $1\,\mathrm{h}^{-1} \mathrm{Mpc}$ of the filament axis, while those in filament outskirts lie between 1 and 2 $\mathrm{h}^{-1} \mathrm{Mpc}$. This figure was generated using our SDSS-DR18 dataset and environment classification scheme.
}
    \label{fig:entornos}
\end{figure*}

As a preliminary step toward understanding how the galaxy–halo connection varies with environment, we first examined the distances between galaxy groups and the critical points of the cosmic web identified by DisPerSE, such as nodes and filaments. While these distances do not define discrete environments, they serve as a first-order proxy for the large-scale structure and allow us to explore how halo occupation properties change as a function of proximity to cosmic web features.

To obtain a more physically motivated and discrete classification of the large-scale environment, we applied the method proposed by \citet{Gal_rraga_Espinosa_2023} and implemented, for example, in \cite{rodriguez2025centralgalaxyalignmentsdependence}, which assigns galaxies to well-defined components of the cosmic web. This approach uses distances to the structures identified by DisPerSE, combined with threshold criteria, to segment the cosmic web into five distinct environments:

\begin{itemize}
    \item \textit{Clusters:} Spherical regions of radius $R_{200} \,[\mathrm{h}^{-1} \mathrm{Mpc}]$ centered on Friends-of-Friends (FoF) halos with masses $M \geq 10^{13}\,M_\mathrm{\odot}\,\mathrm{h}^{-1}$. Galaxies within $R_{200}$ of such halos are considered part of this environment.

    \item \textit{Cluster outskirts:} Spherical shells extending from $R_{200}$ to $3R_{200}$ around massive halos, representing galaxies still under the cluster’s gravitational influence but outside its virialized core.

    \item \textit{Filaments:} Ridges of the density field connecting pairs of nodes, as identified by the DisPerSE skeleton. Galaxies located within $1\,\mathrm{h}^{-1} \mathrm{Mpc}$ of a filament axis are assigned to this environment.

    \item \textit{Filament outskirts:} Regions located between $1$ and $2\,\mathrm{h}^{-1} \mathrm{Mpc}$ from a filament axis, corresponding to galaxies near filaments but not embedded in their central spines.

    \item \textit{Others:} Galaxies not associated with clusters or filaments. These typically reside in low-density regions of the cosmic web and are more isolated from prominent structures.
\end{itemize}

This classification provides a robust and interpretable segmentation of the large-scale environment, which we use in Sect.~\ref{subsec:env_hod} to analyze variations in the HOD.

Figure~\ref{fig:entornos} illustrates the spatial distribution of galaxies across the five environments defined above. The three-dimensional visualization highlights galaxy clusters (red), cluster outskirts (blue), filaments (yellow), filament outskirts (brown), and remaining galaxies not directly associated with these structures. Filamentary features were extracted using the DisPerSE algorithm, and galaxy groups are shown as transparent magenta spheres with radii equal to \(3 \times R_{200}\). Galaxies classified as \textit{Filaments} lie within \(1\,\mathrm{h}^{-1} \mathrm{Mpc}\) of a filament axis, while those in \textit{Filament Outskirts} are located between \(1\) and \(2\,\mathrm{h}^{-1} \mathrm{Mpc}\). This visualization, based on our SDSS-DR18 dataset, offers an intuitive depiction of how galaxies trace the underlying large-scale structure.

\begin{figure*}[h!]
    \centering    \includegraphics[width=1\linewidth]{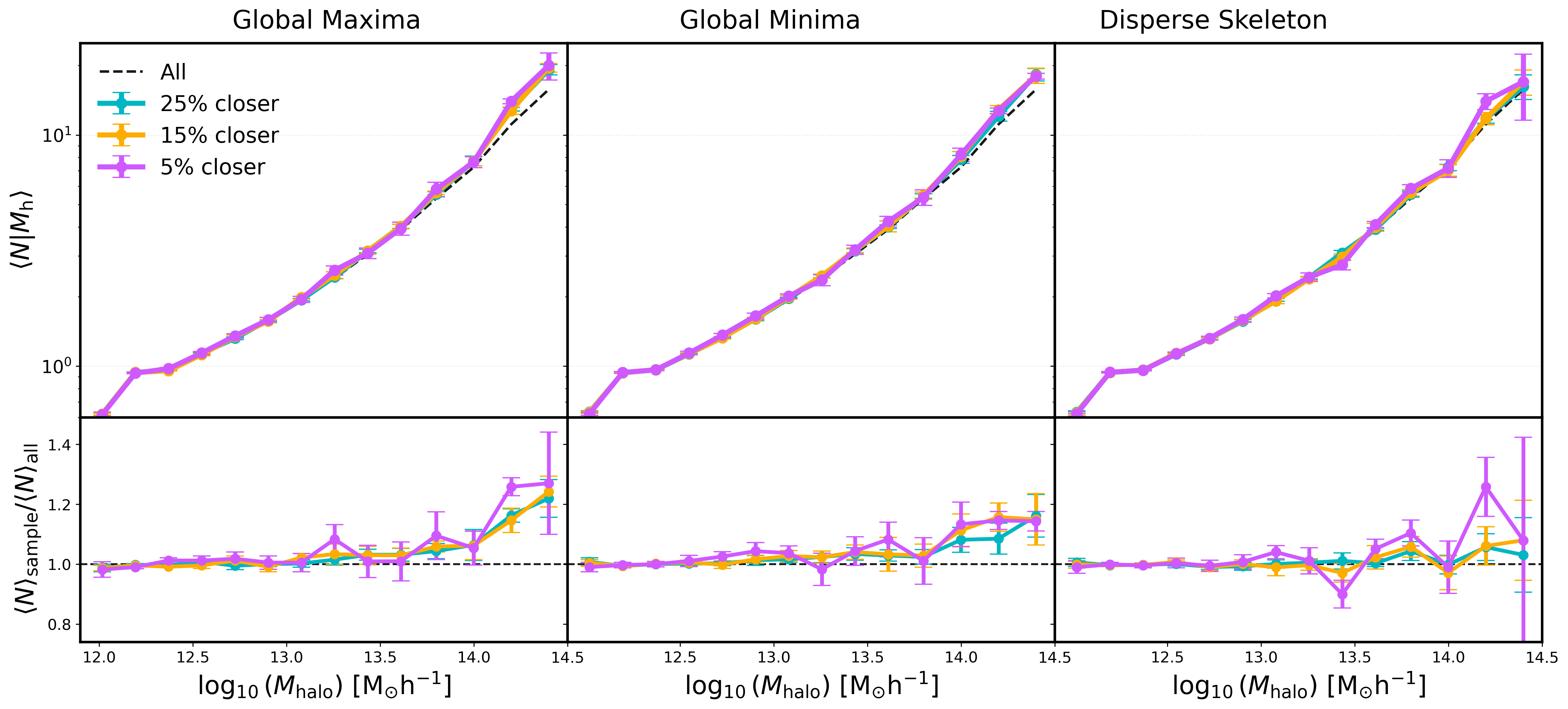}
    \caption{HODs as a function of halo mass for groups at different proximity thresholds. Upper panels: HODs for subsamples within 25\% (cyan), 15\% (orange), and 5\% (violet) of the nearest global maxima (left), global minima (center), and DisPerSE skeleton (right), compared to the complete sample (black dashed line). Lower panels: Ratios of each sample HOD to the complete sample HOD. The larger fluctuations at high masses reflect the smaller number of halos in these regimes. Errorbars represent jackknife uncertainties estimated using 100 spatial subsamples. All results correspond to the absolute magnitude cut $M_{\mathrm{r}} < -20.0$.}

    \label{fig:HODs distances}
\end{figure*}
\section{Exploring Potential Variations of the HOD Across the Cosmic Web}
\label{sec:results}

\subsection{Analysis as a Function of the distance to DisPerSE Critical Points}

Throughout this work, we refer to DisPerSE maxima as Global Maxima, minima as Global Minima, and filaments as the DisPerSE Skeleton. We first investigated whether the HOD correlates with the distance to the nearest such critical point. For this analysis, group subsamples were selected based on the distances measured from the central galaxies of each group to the critical points, with subsequent inclusion of their satellite galaxies. We chose central galaxies to define these distances because they are typically located near the potential minimum of their host halos and are less affected by redshift-space distortions compared to satellites. As a result, their positions provide more stable and physically meaningful references for measuring proximity to large-scale structures. We considered groups within the 25\%, 15\%, and 5\% closest distances to each critical point type, as identified by DisPerSE. Jackknife uncertainties were computed using 100 subsamples, with tests confirming stable variance for $\geq 50$ subsamples. Notably, the error bars increase significantly at the highest mass bins due to the decreasing number of halos in these regimes. The fundamental limitation of this distance-based approach is its inability to isolate pure environments. While a group may be classified as close to a Global Maxima, its central galaxy (and consequently the whole group) might simultaneously reside near a filament or Global Minima, creating ambiguous environmental classifications. This problem originates because DisPerSE identifies only discrete critical points without characterizing their spatial extent or overlaps, as noted in similar analyses by \citet{montero_facu_2024}. Figure~\ref{fig:HODs distances} shows the occupation distributions for the magnitude threshold $M_\mathrm{r} < -20.0$, representative of the behavior observed across all thresholds considered ($M_\mathrm{r} < -21.0$, $-20.5$, $-20.0$, and $-19.76$). We display only this case for clarity, as the trends remain consistent for the other cuts. Across all mass bins, there are no significant HOD variations between the proximity-selected subsamples and the complete sample. Though the larger uncertainties at high masses preclude strong conclusions in these extreme regimes, the HOD ratios (lower panels) fluctuate minimally around unity, confirming that critical point distances alone lack predictive power for halo occupation.

Given these limitations, we proceed with the more rigorous environment classification from Sect.~\ref{sec:environment}, which explicitly accounts for the multidimensional nature of cosmic structures while maintaining the central galaxy-based distance measurements as our fundamental reference.

\subsection{Variations Across Different Environments}
\label{subsec:env_hod}

\begin{figure*}
\centering
\includegraphics[width=1\linewidth]{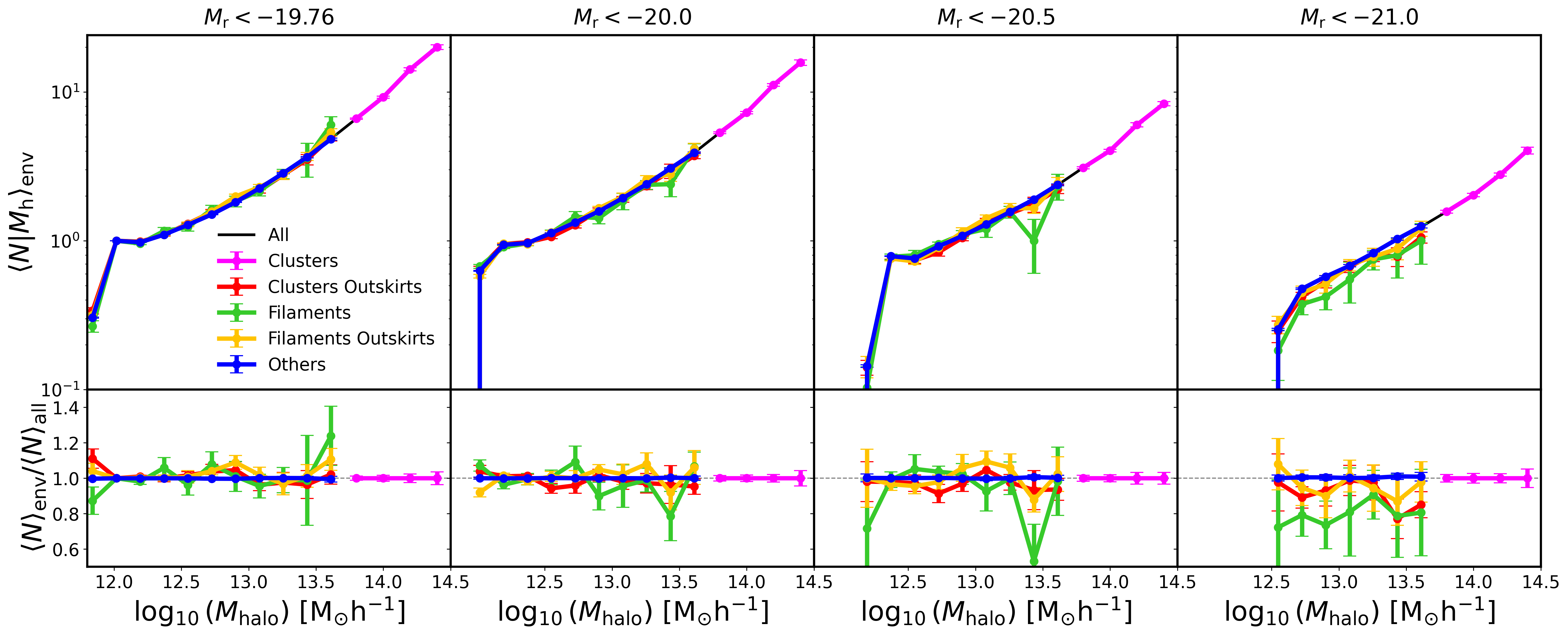}
\caption{
HODs across cosmic environments for magnitude-limited samples ($M_{\mathrm{r}} < -19.76$, $-20.0$, $-20.5$, and $-21.0$). The top panels display mean occupation number versus halo mass for the overall sample (black line) compared to individual environments: Clusters (pink), Clusters Outskirts (red), Filaments (green), Filaments Outskirts (yellow), and Others (blue). Bottom panels show the ratio of each environment's HOD to the overall sample. Errorbars indicate jackknife uncertainties from 100 spatial subsamples.
}
\label{fig:hods_entornos}
\end{figure*}

Using a methodology similar to the one described above, and the environments defined in Sect.~\ref{sec:environment}, we obtained and analyzed the HODs across different environments. As shown in Table~\ref{tab:env_counts}, although the number of groups classified as belonging to the Clusters environment is relatively small, these groups contain a disproportionately large number of galaxies. This is consistent with the definition of clusters as dense, highly populated structures.

\begin{table}[h]
\centering
\caption{Number of groups and galaxies in each environment.}
\begin{tabular}{lrr}
\hline
\textbf{Environment} & \textbf{Groups} & \textbf{Galaxies} \\
\hline
Clusters             & 1,157           & 20,323  \\
Cluster Outskirts   & 2,735           & 4,366   \\
Filaments            & 730             & 1,170   \\
Filament Outskirts  & 2,175           & 3,379   \\
Others               & 110,982         & 169,635 \\
\hline
\end{tabular}
\label{tab:env_counts}
\end{table}

Figure~\ref{fig:hods_entornos} presents the HODs for galaxy groups classified by environment, across all magnitude-limited samples. The top panels show the HODs for each environment, allowing direct comparison of occupation trends, while the bottom panels display the ratio of each environmental HOD to that of the full sample, highlighting relative deviations.

Across the low- and intermediate-luminosity thresholds (the first three in Fig.~\ref{fig:hods_entornos}), the HODs exhibit a high degree of consistency between environments and the full sample. In most cases, the environmental curves closely follow the overall trend (black line), with their ratios remaining near unity and typically fluctuating by less than 10\% over the full halo mass range. This suggests that large-scale environment plays only a limited role in shaping halo occupation. Cluster environments show well-constrained HODs, with uncertainties around 3–5\%. Their outskirts display similarly stable behavior, with slightly larger but still controlled errors. Filamentary environments exhibit greater variation: filament cores show the highest uncertainties, up to 15–20\% in low-occupancy bins, due to low number statistics, while filament outskirts show intermediate scatter, but remain broadly consistent with the global trend within uncertainties. 

A notable exception to the aforementioned trend appears for the brightest galaxy sample ($M_{\mathrm{r}}<-21.0$), where intermediate-mass groups classified in filament cores show a systematic tendency toward lower occupation numbers relative to the full sample (by 15-20 $\%$). Although this trend is limited by statistical uncertainties, it may indicate a reduced efficiency of galaxy formation in these lower-density environments at high luminosities. The environment labeled as Others displays small statistical uncertainties due to its large sample size. However, since it includes all regions not identified as clusters or filaments, it likely encompasses a heterogeneous mix of environments. As such, we refrain from interpreting any physical trends from this category. 

Overall, except for the potential suppression seen in filament cores at the brightest luminosities, the HODs remain fairly similar across environments. These results reinforce the idea that halo mass is the dominant factor governing galaxy occupation, with minimal modulation by large-scale environment within the sensitivity of our current analysis.

\subsection{Variations Across Environments and Color}
\label{subsec:env_color}
 
\begin{figure*}
    \centering
    \includegraphics[width=1\linewidth]{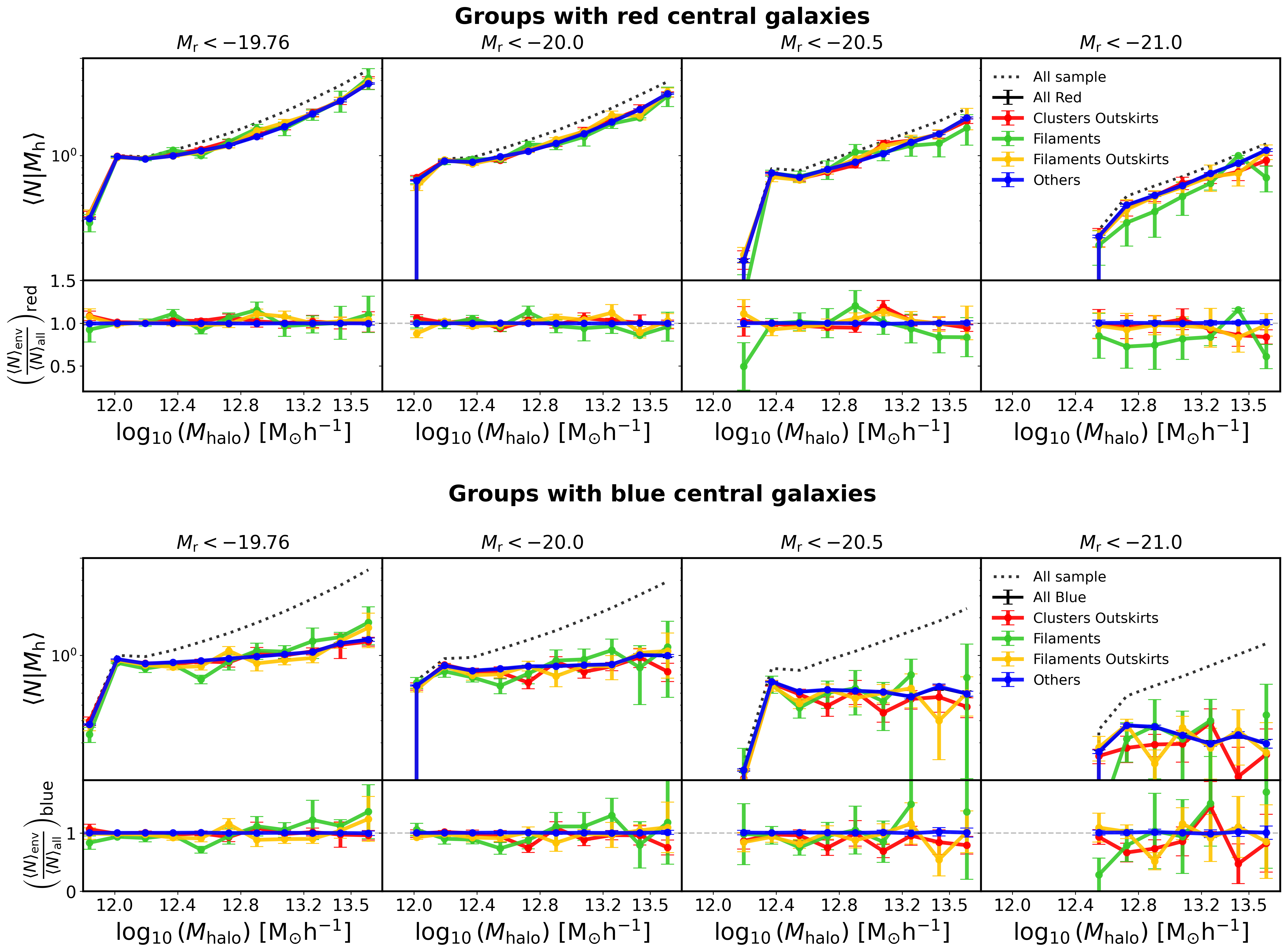}
    \caption{HODs segregated by galaxy color and environment. Top row: HODs (left) and their corresponding ratios (right) for red central galaxies ($(g-r) > 0.83$). Bottom row: Equivalent analysis for blue centrals ($(g-r) \leq 0.83$). Each column represents a different magnitude cut ($M_{\mathrm{r}} < -19.76$, $-20.0$, $-20.5$, $-21.0$). Solid black lines show the complete sample (All), while colored symbols represent environmental subsamples (Clusters, Clusters Outskirts, Filaments, Filaments Outskirts, Others). Ratio panels compare each environment's HOD to the corresponding color-classified complete sample (All red or All blue).}
    \label{fig:redblue}
\end{figure*}

Building on our environmental analysis, we now examine how galaxy color influences the HOD across different environments. We classify central galaxies as red ($(g - r) > 0.83$) or blue ($(g - r) \leq 0.83$) following \cite{rodriguez_artale2022}, while preserving their satellite populations to maintain group integrity in our HOD calculations. This approach allows us to investigate potential correlations between central galaxy properties and their halo occupation patterns in different cosmic environments. Figure \ref{fig:redblue} presents the color-dependent HOD analysis, organized in two distinct rows. The upper row displays the HODs and their corresponding ratios for red central galaxies across all environmental classes, while the lower row shows the equivalent results for blue centrals. Each column represents a different magnitude threshold ($M_{\mathrm{r}} < -19.76$, $-20.0$, $-20.5$, and $-21.0$), with the black line indicating the overall sample and colored symbols marking the environmental subsamples. The ratio panels displays how each environment's occupation compares to the HOD of all galaxies of the same color, maintaining the classification scheme established in Sect.~\ref{subsec:env_hod}. For red central galaxies, the HOD follows the global trend across all environments and halo masses. However, blue central galaxies exhibit a systematically lower number of satellite galaxies, particularly in halos with higher masses. 

In the combined population (Fig.~\ref{fig:hods_entornos}) deviations appeared for the brightest galaxy sample ($M_{\mathrm{r}} < -21.0$), where intermediate-mass groups in filaments showed a systematic tendency toward lower occupation numbers. By separating galaxies into blue and red populations in Fig.~\ref{fig:redblue}, we find that this effect can be mostly attributed to red central galaxies (upper row).

The subpanels of Fig. \ref{fig:redblue} display the ratio of the HOD in each environment to the general population of galaxies of the same color. For both red and blue central galaxies, these ratios remain close to unity, suggesting that the occupation number is consistent across environments. For blue centrals, the HODs show a larger scatter, particularly in environments like Filaments and Others, where the number of satellites is reduced.

\begin{figure*}
    \centering
    \includegraphics[width=1\linewidth]{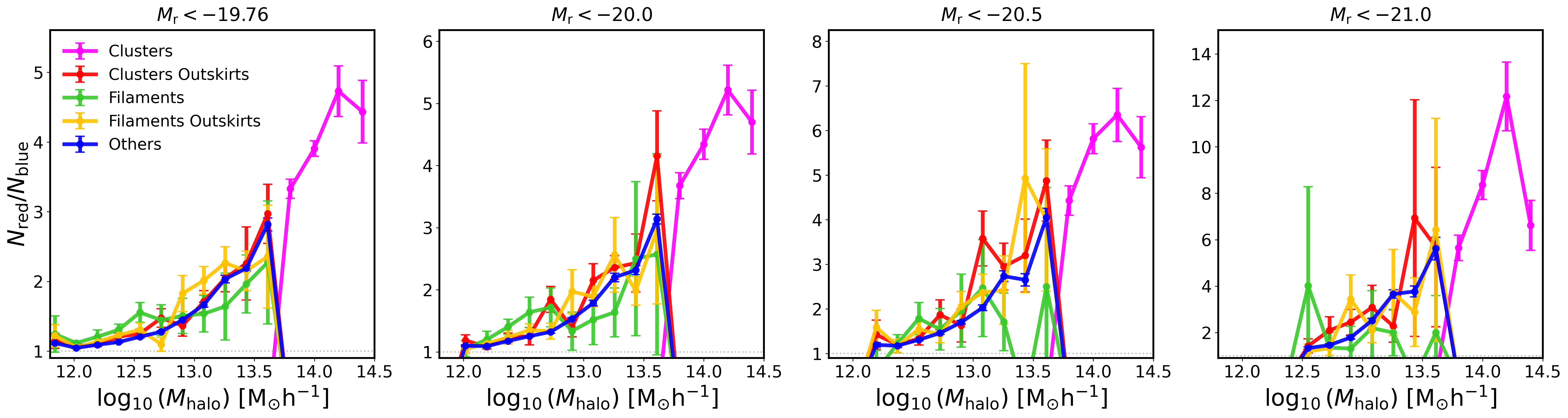}
    \caption{Red-to-blue galaxy ratio ($N_{\mathrm{red}}/N_{\mathrm{blue}}$) as a function of halo mass ($\log_{10}(M_{\mathrm{halo}}/[\mathrm{M_\odot}\mathrm{h}^{-1}]$) for different galactic environments (color-coded) and four magnitude cuts (panels). The ratio increases systematically with halo mass in all environments, with the steepest growth occurring in the densest regions (clusters $>$ cluster outskirts $>$ filaments $>$ filament outskirts $>$ field). This mass- and density-dependent trend persists consistently across all magnitude cuts. Error bars show propagated uncertainties from jackknife-estimated HOD errors.}
    \label{fig:ratio}
\end{figure*}

To further investigate the environmental dependence of satellite galaxy populations, Fig.~\ref{fig:ratio} shows the ratio of satellites hosted by red versus blue central galaxies as a function of halo mass, deparated by environment. Across all environments and magnitude thresholds ($M_{\mathrm{r}} < -21.0$, $M_{\mathrm{r}} < -20.5$, etc.), red centrals host significantly more satellites than blue centrals, with the disparity increasing sharply with halo mass. Fig.~\ref{fig:ratio} once again shows the modest impact of large-scale environment on halo occupation, with the satellite ratio depending primarily on halo mass. As expected, filaments appear to slightly deviate from this general trend, displaying a relatively flat behavior at higher halo masses, although the uncertainties prevent us from drawing any firm conclusions.

These trends align with established observational results on satellite quenching and galactic conformity. \citet{Peng2012} and \citet{wetzel2013} demonstrated that satellites in halos with red centrals are more likely to be quenched, particularly in massive systems, while \citet{Zehavi2011} showed that the HOD is strongly color-dependent, with red centrals hosting more satellites at fixed halo mass. The physical interpretation aligns with recent findings on environmental processing timescales. Specifically, \citet{oxland2024} showed that in dense environments, star formation is quenched more rapidly than morphological transformation, with pre-processing mainly impacting star formation activity. This supports our scenario where red centrals—typically residing in massive halos and formed through mergers—are more effective at retaining and quenching satellites. Conversely, blue centrals, generally at earlier evolutionary stages with ongoing star formation and active feedback, tend to retain fewer satellites. The accelerated quenching in dense environments further explains why the difference in satellite fractions between red and blue centrals is most pronounced in clusters, where environmental effects are strongest.


\subsection{Robustness tests}

To ensure the reliability of our conclusions, we assessed the robustness of the results against variations in the methodological choices, particularly those related to the identification of the cosmic web. Since our environmental classification relies on structures detected by DisPerSE, it is crucial to understand how sensitive our findings are to the algorithm’s internal parameters. By systematically varying these parameters, we can evaluate the stability of the derived environmental trends and verify that our main conclusions are not artifacts of a specific configuration. In this section, we explore how DisPerSE's parameters, specifically persistence and smoothing (see Subsection~\ref{subsec:disperse}), influence the detection and characterization of structures within the cosmic web. These parameters control the statistical significance and geometric smoothing of the identified features, enabling a tailored analysis of the cosmic web.

Table \ref{tab:critical points} summarizes the number of critical points identified in the Main Galaxy Sample (MGS) of SDSS for various persistence thresholds (\(3\sigma\) and \(5\sigma\)) and levels of smoothing (no smoothing, one smoothing, and two smoothings). As the persistence threshold and smoothing level increase, the number of critical points detected decreases significantly. This behavior arises because higher persistence thresholds filter out less significant structures, while increased smoothing eliminates finer details, leaving only the most prominent features. 

\begin{figure*}
    \centering    \includegraphics[width=1\linewidth]{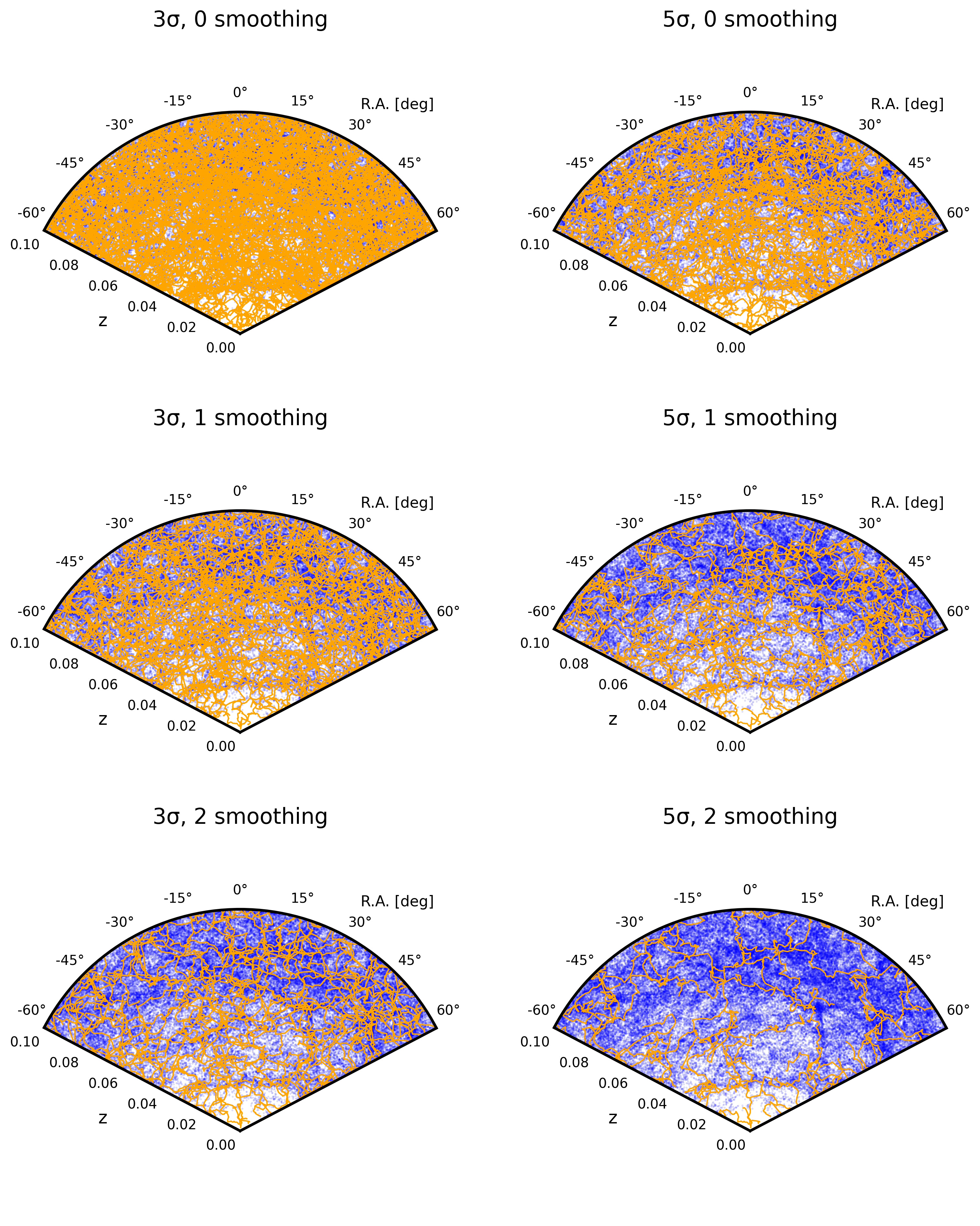}
    \caption{Distribution of central galaxies (blue dots) and cosmic‐web filaments (orange lines) 
    in polar coordinates, where the angular axis represents Right Ascension (RA) and the radial 
    axis corresponds to redshift. Each of the six panels shows filaments extracted with different 
    combinations of persistence threshold (3\,$\sigma$ or 5\,$\sigma$) and smoothing level 
    (0, 1, or 2) using DisPerSE. The underlying galaxy sample is identical across all panels 
    to highlight the effect of varying DisPerSE parameters on filament detection.}
    \label{fig:filamentos_disperse}
\end{figure*}

\begin{table}[h!]
\centering
\caption{Number of filament vertices at redshift $\mathrm{z} < 0.1$ by persistence level.}
\begin{tabular}{lcc}
\hline
\textbf{Persistence} & \textbf{3$\sigma$} & \textbf{5$\sigma$} \\
\hline
0 & 47\,978 & 22\,863 \\
1 & 17\,133 & 8\,490 \\
2 & 8\,238 & 3\,558 \\
\hline
\end{tabular}
\label{tab:critical points}
\end{table}

Figure \ref{fig:filamentos_disperse} illustrates the distribution of filaments identified by DisPerSE for the same persistence and smoothing parameters. The panels clearly demonstrate how these settings impact the complexity and prominence of the detected filamentary structures.

In the upper-left panel (\(3\sigma\), no smoothing), the number of filaments is significantly higher, indicating the inclusion of many smaller and less prominent structures. This is a consequence of the low persistence threshold, which allows DisPerSE to detect a wide range of features, including noise and minor filaments. The absence of smoothing preserves fine details, resulting in a dense and intricate web-like structure.

\begin{figure*}[h!]
    \centering
    \includegraphics[width=1\linewidth]{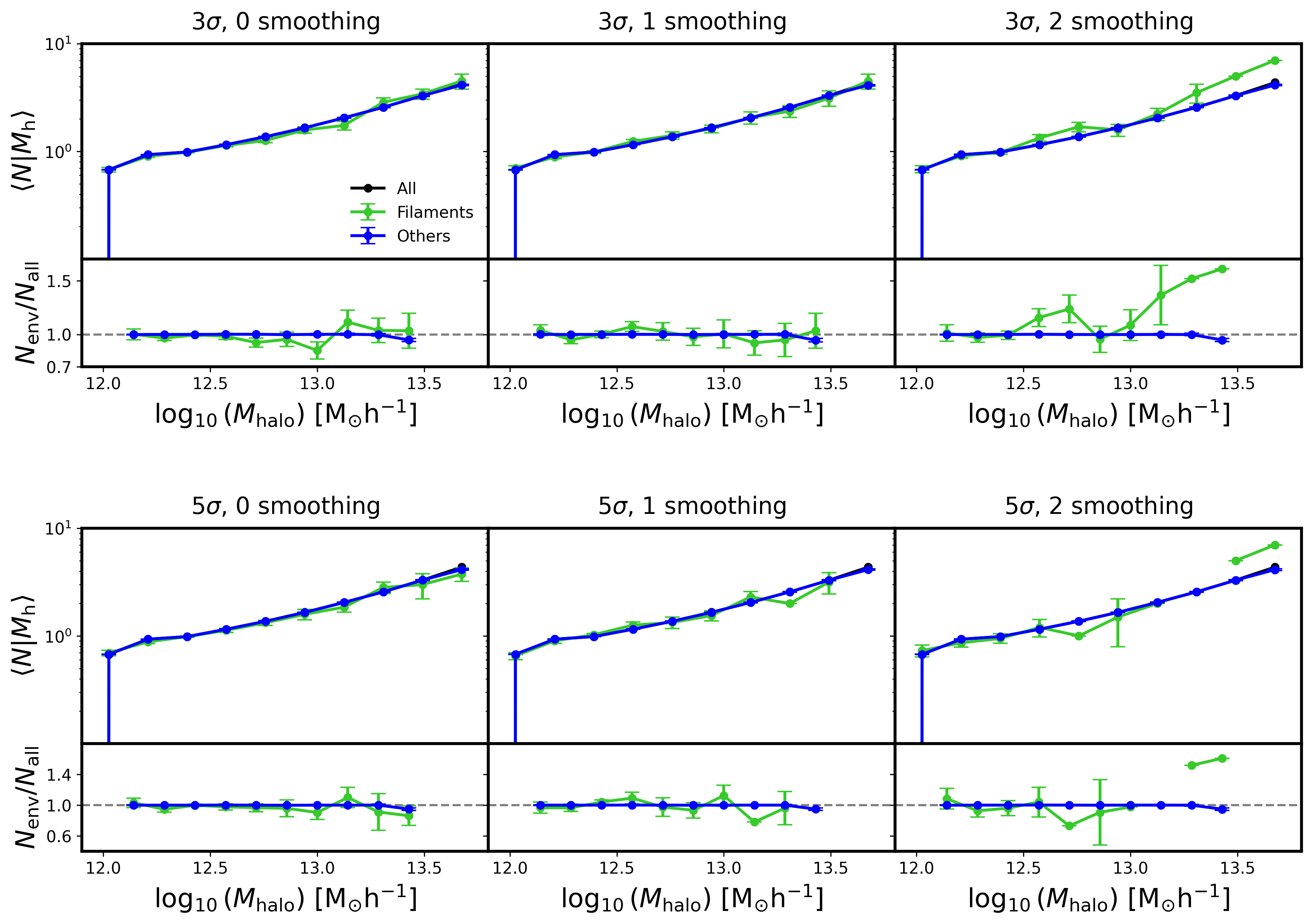}
    \caption{HODs and corresponding ratios between environmental and global HODs for a magnitude cut of $M_{\mathrm{r}} < -20.0$. The top two rows show the results using a DisPerSE persistence threshold of $3\sigma$, while the bottom two rows correspond to $5\sigma$. Each column indicates a different smoothing level (0, 1, and 2). The top and fourth rows show the mean number of galaxies as a function of halo mass, $\langle N | M_{\mathrm{h}} \rangle$, for all halos (black) and for those in filaments (yellow), filament outskirts (orange), and other regions (red). The second and fifth rows show the ratio of the environmental HODs to the global one, highlighting relative variations across halo masses and environments.}
    \label{fig:hod_parameters}
\end{figure*}

As smoothing is increased (top-middle and top-right panels), the finer structures are progressively removed, leaving only the larger-scale filaments. This illustrates the effect of smoothing in simplifying the cosmic web by suppressing small-scale noise and enhancing the coherence of prominent features.

The bottom row (\(5\sigma\)) reflects the effect of a higher persistence threshold. Compared to the \(3\sigma\) panels, the number of filaments decreases substantially as only statistically significant structures are retained. In the absence of smoothing (bottom-left panel), the detected filaments are less dense but still exhibit substantial complexity. As smoothing levels increase (bottom-middle and bottom-right panels), the filamentary network is further simplified, with only the most significant structures remaining. The bottom-right panel (\(5\sigma\), 2 smoothing) represents the most simplified version of the cosmic web, where only the strongest and most coherent filaments are preserved.

This analysis highlights how DisPerSE's parameters can be tuned to study different aspects of the cosmic web. Lower persistence thresholds and minimal smoothing are useful for exploring the detailed structure of the web, while higher thresholds and greater smoothing emphasize the most prominent and statistically significant features.

The analysis of Fig. \ref{fig:hod_parameters} shows the HOD for two filament-associated environments, Filaments and Filaments Outskirts, as a function of the halo mass. The upper panels present the mean occupation number, \(\langle N | M_{\mathrm{h}} \rangle\), for the overall population (black line) and for specific combinations of persistence parameters (\(3\sigma\) and \(5\sigma\)) and smoothing levels (0, 1, and 2).

In Filaments, shown in the left panels, the HOD shows that, for a low persistence threshold (\(3\sigma\)), less prominent structures are identified, resulting in more scattered trends, particularly as the smoothing level increases. In contrast, for \(5\sigma\), the HOD becomes smoother and aligns more closely with the general trend, since only the most significant structures are considered. This behavior is consistent in the Filaments Outskirts environment, where the differences between parameter configurations are similar.

The bottom panels show the HOD ratios for each configuration relative to the overall population. In both environments, the ratios remain close to unity, indicating that variations in persistence and smoothing have a limited impact on the average occupation for most halo mass bins. However, at \(3\sigma\), the ratios exhibit greater fluctuations, particularly in low-mass halos (\(M_{\mathrm{h}} < 10^{12.5} M_{\mathrm{\odot}} \mathrm{h}^{-1}\)), reflecting the inclusion of less prominent structures. At \(5\sigma\), the ratios are more stable, demonstrating that focusing on more significant structures yields more robust and consistent results.

In both environments, general trends indicate that higher mass halos (\(M_{\mathrm{h}} > 10^{13} M_{\mathrm{\odot}} \mathrm{h}^{-1}\)) exhibit similar occupations regardless of the DisPerSE parameters, while in lower mass halos, configurations with lower persistence and smoothing detect greater variations. This reinforces the idea that DisPerSE parameters allow for tailoring the analysis of the cosmic web, capturing small and detailed structures in low-persistence configurations and emphasizing the most prominent and stable structures in high-persistence and smoothing configurations.


\section{Discussion and Conclusions}
\label{sec:conclusions}

This work investigates how galaxies occupy dark matter halos across the cosmic web,  focusing on the impact of different environments and galaxy properties on the Halo Occupation Distribution (HOD). Using observational data from the SDSS DR18, we rely on a publicly available group catalogue in which galaxy groups, which are used as proxies for dark matter halos, have been previously identified, with assigned halo masses, group memberships, and unique identifiers \citep{Rodriguez2020}. To characterize the large-scale structure, we use an additional publicly available catalogue based on the DisPerSE algorithm, which traces the cosmic web and identifies its critical points \citep{Malavasi2020}. We then examine how environmental classifications and intrinsic galaxy properties influence the HOD across distinct cosmic environments.

A key finding of our analysis is the ineffectiveness of simple distance-based environmental classifications -- constructed from the DisPerSE critical points -- in distinguishing HOD differences. This straightforward approach fails to capture the complex and overlapping nature of cosmic environments, such as the coexistence of filaments, voids, and clusters. In practice, what this method primarily reveals is the well-known tendency of galaxies to lie near clusters and avoid voids, rather than providing meaningful insight into how HOD varies with environment. As a result, separating galaxies into distinct environments is essential for a more informative analysis of HOD variations.

To improve upon simple distance-based classifications, our HOD analysis incorporated discrete environmental categories, including clusters, cluster outskirts, filaments, filament outskirts, and a broader category encompassing all other regions. Even with this more refined approach, no significant differences are observed across these environments, reinforcing the overall invariance of group occupation throughout the cosmic web. The only notable deviation is found for intermediate-mass groups, which appear to host 15–20$\%$ fewer luminous galaxies when located in filaments, although this difference remains within the uncertainties of the measurement. The stability of the HOD suggests that, within the limits of our analysis, the large-scale environment has little effect on halo occupation, especially when intrinsic galaxy properties are not accounted for. These results support the view that environmental effects on the HOD are subtle and may only become apparent in extreme environments and when the parameter space is expanded to include additional galaxy characteristics (e.g., \citealt{Alfaro2020, Alfaro2021, Alfaro_2022}).

As expected, when galaxies are classified by color, clear differences emerge in the HODs. Red central galaxies consistently host significantly more satellite galaxies than blue central galaxies. Our analysis confirms this well-established trend across all environments and volumes (e.g., \citealt{Zehavi2011,Peng2012,wetzel2013,Campbell2015,oxland2024}). This result highlights the strong dependence of the HOD on the properties of the central galaxy, with galaxy color serving as a reliable indicator of satellite populations. Blue central galaxies, indicative of younger, actively star-forming systems, tend to host far fewer satellites than red central galaxies, which are more evolved and quiescent. Once again, slight deviations are mainly observed in filaments for the brightest galaxy samples, with our results suggesting that this effect may be relevant only for red galaxies. These subtle differences hint at the potential of combining environmental and intrinsic galaxy properties, especially with improving spectroscopic data in upcoming years. 

  
We have also verified that the characterization of filaments is highly sensitive to the choice of DisPerSE parameters. In particular, the smoothing scale has a significant impact on both the identification of filaments and the subsequent environmental classification of galaxies. While these variations have little effect on the resulting HODs across most environments, once again, notable differences arise in filament regions, where satellite statistics can vary by up to 50$\%$ in certain mass ranges when the largest smoothing scale is adopted. Although still subtle, these variations warrant further investigation to better understand the connection between HOD and filamentary structure.



The consistency of the HOD that we measure observationally is in fairly good agreement with previous results, although some differences exist. Using the Illustris-TNG hydrodynamical simulations (\citealt{Pillepich2018b,Pillepich2018,Springel2018}) and DisPerSE, \cite{Perez2024} found that the HOD for halos in filaments was remarkably similar to that of the total galaxy sample, suggesting halo mass is the dominant factor in galaxy occupation.  In the SDSS, \cite{Alfaro_2022} discovered HOD variations but only in extreme environments, such as very underdense voids and FVSs. Moving forward, it would be valuable to extend our analysis to lower masses, refine environmental classification schemes, and explore also alternative cosmic web definitions. 

A relevant connection with our results lies in the context of occupancy variations — the secondary dependencies of the HOD on internal halo properties beyond halo mass. This effect is one of the main interpretations of the possible manifestations of halo assembly bias in the galaxy population (\citealt{Zehavi2018}). In this view, galaxy assembly bias may result from the underlying halo assembly bias, combined with a dependence of halo occupancy on the same secondary halo property at fixed mass. Although such trends have been measured in hydrodynamical simulations at high statistical significance \citep{Artale2018}, definitive observational confirmation remains elusive \citep{Salcedo2022, Wang2022, Alam2024}. At face value, the consistency of the HOD across the cosmic web in the SDSS, as reported in this work, appears to leave little room for occupancy variations. It is reasonable to expect that the strongest secondary effect — beyond halo mass — would stem from the environment (which is indeed the case for large-scale bias; see, e.g., \citealt{Musso2018, Borzyszkowski2017,Paranjape2018, montero_facu_2024}). 

This work is particularly timely, as new data from upcoming surveys, such as the Dark Energy Spectroscopic Instrument (DESI; \citealt{DESI2016}), Prime Focus Spectrograph (PFS; \citealt{Tamura2016}), Euclid \citep{Laureijs2011}, and the Vera C. Rubin Observatory \citep{Ivezic2019}, will soon become available. These data will enable a more comprehensive understanding of how galaxies and groups interact with the cosmic web and its underlying structures.

\begin{acknowledgements}
      FR thanks the support by Agencia Nacional de Promoci\'on Científica y Tecnológica, the Consejo Nacional de Investigaciones Científicas y Técnicas (CONICET, Argentina) and the Secretaría de Ciencia y Tecnología de la Universidad Nacional de Córdoba (SeCyT-UNC, Argentina). ADMD acknowledges support from the Universidad Técnica Federico Santa María through the Proyecto Interno Regular \texttt{PI\_LIR\_25\_04}. FR and ADMD thank the ICTP for their hospitality and financial support through the Junior Associates Programme 2023–2028 and Regular Associates Programme 2022–2027, respectively.
\end{acknowledgements}

%
%

\bibliographystyle{aa} 
\bibliography{references} 

\begin{appendix} 

\end{appendix} 

\end{document}